\documentclass[a4paper,11pt]{article}

\usepackage{amsmath,amssymb,color,graphics,amscd,amsfonts,epsf}
\allowdisplaybreaks[4]

\date{}
\begin{document}

\begin{flushright} 
 
WIS/06/09-MAY-DPP\\
APCTP Pre2009 - 005

\end{flushright} 

\vspace{0.1cm}

\begin{center}
  {\LARGE
  
   Four-dimensional ${\cal N}=1$ super Yang-Mills  
   from matrix model
   
  }
\end{center}
\vspace{0.1cm}
\vspace{0.1cm}
\begin{center}

         Masanori H{\sc anada}$^{a}$\footnote
         {
E-mail address : masanori.hanada@weizmann.ac.il},  
         Lorenzo M{\sc annelli}$^{a}$\footnote
         {
E-mail address : lorenzo.mannelli@weizmann.ac.il} 
and 
         Yoshinori M{\sc atsuo}$^{b}$\footnote
         {
E-mail address : ymatsuo@apctp.org}         

\vspace{0.3cm}

${}^{a}$
{\it Department of Particle Physics, Weizmann Institute of Science,\\
     Rehovot 76100, Israel }\\

${}^{b}$
{\it Asia Pacific Center for Theoretical Physics,\\
Pohang, Gyeongbuk 790-784, Korea}\\
\end{center}


\vspace{1.5cm}

\begin{center}
  {\bf abstract}
\end{center}

We consider a  supersymmetric matrix quantum mechanics.   
This is obtained by adding 
Myers and mass terms to the dimensional reduction of 4d
${\cal N}=1$ super Yang-Mills theory to one dimension. 
Using this model we construct 4d ${\cal N}=1$ super Yang-Mills theory
in the planar limit by using the Eguchi-Kawai equivalence. 
This regularization turns out to be free from the sign problem 
at the regularized level. 
The same matrix quantum mechanics is also used to provide a nonperturbative formulation of 4d ${\cal N}=1$ 
super Yang-Mills theory on a noncommutative space.

\newpage
\newpage

\section{Introduction}
\hspace{0.51cm}

Supersymmetry  
is a promising framework for
physics beyond the standard model.   
For this reason it is important to understand its nonperturbative aspects such as 
confinement and the mechanism of supersymmetry breaking.  
Usually lattice regularization provides a tool to study field theories in the nonperturbative regime.
However, a technical obstacle arises in this type of regularization, namely it is difficult to keep supersymmetry
(although some progress has been achieved for some specific kind of theories,
for a review see \cite{CKU09}). 
In order to avoid this obstacle for large-$N$ supersymmetric Yang-Mills theories (SYM),
we can use another regularization method known as Eguchi-Kawai reduction 
\cite{EK82} .

Another motivation to study SYM theories in the large-$N$ limit
is that they are expected to describe 
the nonperturbative dynamics of string theory 
\cite{BFSS96,IKKT96,DVV97,Maldacena97,IMSY98}.  
For instance, $(0+1)$-dimensional 
maximally supersymmetric $U(N)$ gauge theory 
is conjectured to be dual to type IIA superstring theory 
on a black 0-brane background \cite{IMSY98}. 
This specific example has been studied 
using Monte-Carlo simulation \cite{HNT07,CW07,HNT08}; 
using these numerical techniques the stringy $\alpha'$ corrections 
can be evaluated \cite{HNT08}. 
There have been much efforts to study the $0+0$-dimensional theory \cite{IKKT96} numerically, too. 
See e.g. \cite{KNS98,AABHN00_1,AABHN00_2,mean_field}. 
Finally, large-$N$ Yang-Mills theories are interesting on their own because 
they might be solvable analytically \cite{KKN98} 
while preserving essential features of 
QCD with $N=3$ (for a recent review, see \cite{Teper08}).  

As previously mentioned, the Eguchi-Kawai equivalence \cite{EK82} 
can be used as an alternative method to regularize large-$N$ SYM. 
The main idea of this method is that large-$N$ gauge theories 
are equivalent to certain lower dimensional matrix models. 
Furthermore in this prescription the degrees of freedom of the reduced spaces are 
embedded in the infinitely large matrices.
A UV regularization can be introduced
by taking the size of the matrices to be large but finite. 
This regularization, differently from the lattice one, does not break supersymmetry. 
With such a motivation, 
in \cite{IIST08} a nonperturbative formulation of 
the maximally supersymmetric Yang-Mills in four dimensions was proposed. 
The authors of \cite{IIST08} have considered a particular solution of 
the BMN matrix model \cite{BMN02},  
namely a set of concentric fuzzy spheres, 
which has been argued to be stable due to its BPS nature. 
Expanding the BMN matrix model about this background, 
the 4d ${\cal N}=4$ SYM was recovered
through the Eguchi-Kawai equivalence.  

In this paper, 
we provide a nonperturbative formulation of 4d ${\cal N}=1$ 
(pure) SYM in the planar limit by using the technique introduced in \cite{IIST08}. 
There are two main motivations to extend the results presented in \cite{IIST08}.
Firstly, 4d ${\cal N}=1$ supersymmetric theories are 
more interesting as a candidate of new physics in the LHC, 
and it is important to 
consider the $\mathcal N=1$ (pure) SYM as a simplest example. 
4d $\mathcal N=1$ SYM is dynamically richer than 
4d ${\cal N}=4$ and given that 
there is no known gravity dual of 4d ${\cal N}=1$ SYM 
providing analytical results, 
numerical simulations are a valuable tool. 
Even though in principle 4d ${\cal N}=1$ SYM 
on the lattice can be studied without fine tuning\footnote{
In the context of lattice regularization, fine tuning means adding 
counterterms in order to restore supersymmetry in the continuum limit.}
it is computationally very demanding, 
and a detailed study is difficult
(for recent numerical studies see 
\cite{4dN=1LatticeSimulation}). 
On the contrary, our supersymmetric matrix models 
would require less resources, and allow a better numerical analysis  of ${\cal N}=1$ SYM.

Secondly, several groups are seriously studying 4d ${\cal N}=1$ SYM on the lattice, 
using conventional computationally demanding numerical techniques. 
When the results of these studies become available they 
could be used to further check the validity of the Eguchi-Kawai regularization. 
After its validity has been further confirmed, the Eguchi-Kawai construction 
could be used to analyze field theories with extended supersymmetries \cite{IIST08}, 
which cannot be studied by using the lattice unless we introduce fine tuning. 

In the first part of this paper we formulate the Eguchi-Kawai reduction of 
4d ${\cal N}=1$ SYM by using a BMN-like mass-deformed matrix quantum mechanics 
with four supersymmetries. 
In the regularization of 4d ${\cal N}=4$ SYM introduced by \cite{IIST08}, 
only 16 out of 32 supercharges are kept unbroken and
the restoration of the other 16 supersymmetries is not obvious, 
although supporting evidence has been found in \cite{IIST08,IKNT08}. 
In the present case, all 4 supercharges are manifestly kept unbroken 
and hence we expect that 4d ${\cal N}=1$ SYM is recovered in the continuum limit. 

In the second part of this paper we consider 
noncommutative super Yang-Mills theories in four dimensions. 
These appear, for example, in string theory 
as effective theories on D-branes with flux. 
Given that noncommutative Yang-Mills theories have a D-brane origin,  
they can be regularized by using matrix models as explained in  \cite{AIIKKT99}. 
Super Yang-Mills in flat noncommutative space is obtained by studying the theory 
in a background satisfying the Heisenberg algebra $[\hat{x},\hat{y}]=i\theta$, 
which cannot be realized using finite-$N$ matrices.
It turns out that the Heisenberg algebra can be described at finite-$N$ level 
by considering compact fuzzy manifolds like the fuzzy sphere embedded in flat space. 
The flat noncommutative space is then recovered as the tangent space to these fuzzy manifolds.
One unsatisfactory property of this prescription is that 
transverse directions are necessary for embedding the compact fuzzy spaces 
into flat space (for example, if we embed $S^2$ into ${\mathbb R}^3$ 
there is one transverse direction). 
In the field theory description these directions turn into scalar fields, 
and therefore, only noncommutative gauge theories with scalars can be realized in this way. 
In the case of supersymmetric models, this implies that it is only possible to regularize 
theories with extended supersymmetries.%
\footnote{If we use the twisted Eguchi-Kawai model \cite{GAO82}, 
which is written in terms of unitary matrices, 
we do not need transverse directions. 
However it is difficult to supersymmetrize it.} 
In this article we show that by using our matrix model 
with the background proposed in \cite{IIST08} in an appropriate limit, 
we can regularize 4d ${\cal N}=1$ noncommutative 
{\it pure} super Yang-Mills. 
Using this construction, the transverse direction becomes
an ordinary {\it commutative} coordinate and as a consequence 
we recover pure ${\cal N}=1$ SYM with no additional scalars.

This paper is organized as follows.  
In \S\ref{section:EguchiKawaiModel} we review the Eguchi-Kawai equivalence. 
In \S\ref{sec:QEK} we explain its deformation, 
namely the ``quenched'' Eguchi-Kawai model, which we then use to formulate 4d ${\cal N}=1$ SYM. 
In \S\ref{EK model for 4d N=1} we provide a 
supersymmetric matrix quantum mechanics, and applying the method explained in \S\ref{sec:QEK},
we provide the Eguchi-Kawai formulation of 4d ${\cal N}=1$ SYM. 
In \S~\ref{sec:sign problem} we prove that this regularization does not suffer from the sign problem. 
In \S\ref{sec:NCSYM} we provide a nonperturbative formulation of 
4d ${\cal N}=1$ SYM on noncommutative space. 
In Appendix~\ref{sec:4dN=1} we introduce four-dimensional ${\cal N}=1$ SYM 
on the three sphere and express it in a form which is convenient for our purpose. 

\section{The Eguchi-Kawai reduction}\label{section:EguchiKawaiModel}
\hspace{0.51cm}
In this section we review the Eguchi-Kawai equivalence \cite{EK82}. 
The equivalence guarantees that $D$-dimensional $SU(N)$ gauge theory and 
its one-point reduction are equivalent if the global $(Z_N)^D$ symmetry 
of the latter is not broken. 
In the bosonic case, however, 
this symmetry is broken for $D>2$ in the 0d theory; it is unbroken 
only above a critical volume \cite{NN03}. 
To cure this problem, deformations of the 0d theory, 
the quenched \cite{BHN82,Parisi82} and twisted \cite{GAO82} Eguchi-Kawai models 
(QEK,TEK) were proposed soon after the original one.%
\footnote{
Another recent proposal can be found in \cite{UY08}. 
}
In these models, deformations are introduced such that 
$\left(Z_N\right)^D$-unbroken backgrounds become stable. 
However, recently it was found that 
both TEK \cite{AHHI07,TV06,BNSV06} and QEK \cite{BS08} 
fail at very large-$N$ -- their deformations
cannot stabilize the backgrounds completely. 
On the other hand, by combining quenched and/or 
twisted prescriptions with supersymmetry, 
the background can be stabilized \cite{AHHI07,AHH08}. 

In \S\ref{sec:QEK} we review the diagrammatic approach to 
the quenched Eguchi-Kawai model (QEK) \cite{Parisi82}. 
First we consider the simplest case, namely the equivalence between 
matrix quantum mechanics and the zero-dimensional matrix model, 
and then we proceed with the Eguchi-Kawai construction of the field theory 
on $S^3$ \cite{IIST08}. 
\subsection{Quenched Eguchi-Kawai model}\label{sec:QEK}
\hspace{0.51cm}
We consider a matrix quantum mechanics with a mass term, 
\begin{eqnarray}
S_{1d}=N\int dt Tr\left(
\frac{1}{2}(D_t X_i)^2
-
\frac{1}{4}[X_i,X_j]^2
+
\frac{m^2}{2} X_i^2
\right),  
\end{eqnarray}
where $X_i\ (i=1,2,\cdots,d)$ are $N\times N$ traceless Hermitian matrices. 
The covariant derivative $D_t$ is 
given by 
$
D_t X_i
=
\partial_t X_i - i[A,X_i]$.  
At large-$N$, this model can be reproduced starting from the zero-dimensional model
\begin{eqnarray}
S_{0d}
=
\frac{2\pi}{\Lambda}
\cdot N
Tr\left(
-
\frac{1}{2}[Y,X_i]^2
-
\frac{1}{4}[X_i,X_j]^2
+
\frac{m^2}{2} X_i^2
\right), 
\end{eqnarray}
where $Y$ and $X_i$ are $N\times N$ traceless Hermitian matrices. 
We embed the (regularized) translation generator into 
the matrix $Y$, 
\begin{eqnarray}
Y^{b.g.}=diag(p_1,\cdots,p_N), 
\qquad
p_k=\frac{\Lambda}{N}\left(
k-\frac{N}{2}
\right). 
\end{eqnarray}
By expanding $Y$ around this background,  
$
Y
=
Y^{b.g.}+A$,   
the Feynman rules of the one-dimensional theory are reproduced at large-$N$, 
as we will see in the following.  

\begin{figure}[tbp]
\begin{center}
\scalebox{0.5}{
\rotatebox{0}{
\includegraphics*{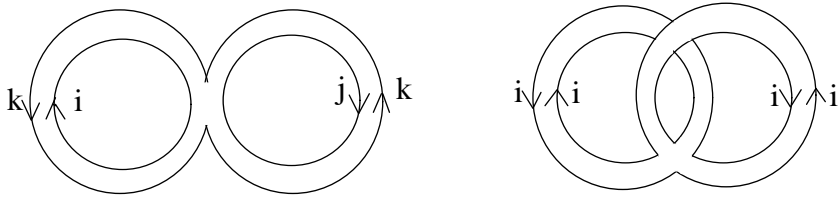}
}}
\caption{Two-loop planar and nonplanar diagrams with quartic interaction vertex. 
}\label{fig:TwoLoop}
\end{center}
\end{figure}

The action can be rewritten as 
\begin{eqnarray}
S_{0d}
=
\frac{2\pi}{\Lambda}
\cdot N
\Biggl\{
\frac{1}{2}
\sum_{i,j}\left|
(p_i-p_j)\left(X_k\right)_{ij}
-
i[A,X_k]_{ij}
\right|^2
+
Tr\left(
-
\frac{1}{4}[X_i,X_j]^2
+
\frac{m^2}{2} X_i^2
\right)
\Biggl\}. 
\end{eqnarray}
We add to it the gauge-fixing and Faddeev-Popov terms 
$
\frac{2\pi}{\Lambda}
\cdot N
Tr\left(
\frac{1}{2}[Y^{b.g.}, A]^2
-
[Y^{b.g.},b][Y,c]
\right)$.  
Then, the planar diagrams are the same as in the 1d theory up to a normalization factor. 
For example, a scalar two-loop planar diagram with a quartic interaction 
(see Fig.\ref{fig:TwoLoop}) is 
\begin{eqnarray}
\lefteqn{
\frac{d(d-1)}{2}
\left(
\frac{1}{2}\cdot\frac{2\pi N}{\Lambda}
\right)
\sum_{i,j,k=1}^N
\frac{(\Lambda/2\pi N)}{(p_i-p_k)^2+m^2}
\frac{(\Lambda/2\pi N)}{(p_j-p_k)^2+m^2}
}\nonumber\\
&\simeq&
\frac{d(d-1)}{4}\cdot\frac{2\pi}{\Lambda}\cdot
N^2
\int^{\Lambda/2}_{-\Lambda/2} \frac{dp}{2\pi}
\int^{\Lambda/2}_{-\Lambda/2} \frac{dq}{2\pi}
\frac{1}{(p^2+m^2)(q^2+m^2)}.   
\end{eqnarray}
The essence of this expression is that {\it the adjoint action of 
the background matrix can be identified with the derivative} 
and
{\it the matrix elements of the fluctuations can be identified with the Fourier modes 
in momentum space}. 
The corresponding diagram in the 1d theory is 
\begin{eqnarray}
\frac{d(d-1)}{4}\cdot Vol
\cdot N^2
\int^{\Lambda/2}_{-\Lambda/2} \frac{dp}{2\pi}
\int^{\Lambda/2}_{-\Lambda/2} \frac{dq}{2\pi}
\frac{1}{(p^2+m^2)(q^2+m^2)}, 
\end{eqnarray}
where $Vol$ is the spacetime volume, and 
hence by interpreting $\Lambda$ and $\Lambda/N$ to be UV and IR cutoffs, 
those diagrams agree up to the factor $\frac{\Lambda}{2\pi}\cdot Vol$. 
The other planar diagrams also correspond up to the same factor. 

The nonplanar diagrams do not have such a correspondence, but in an appropriate limit 
they are negligible. In the 1d theory, by taking a planar limit 
they are suppressed by a factor $N^{-2}$. In the reduced model, they are suppressed 
if the IR cutoff $\Lambda/N$ goes to zero. To see this, let us calculate the 
two-loop nonplanar diagram in Fig.\ref{fig:TwoLoop}, for example. 
It reads 
$
-\frac{d(d-1)}{4m^4}\frac{\Lambda}{2\pi}$, 
which is suppressed by a factor $(\Lambda/N)^2$ compared with planar diagrams. 

Therefore, by taking the limit 
\begin{eqnarray}
N\to\infty, 
\qquad
\Lambda\to\infty, 
\qquad
\frac{\Lambda}{N}\to 0
\end{eqnarray}
the 1d model on ${\mathbb R}$ is reproduced from the 0d model. 
\subsection{Eguchi-Kawai construction of Yang-Mills on $S^3$}\label{sec:3d example}
\hspace{0.51cm}
Next let us construct the Yang-Mills theory on three-sphere 
by using the Eguchi-Kawai equivalence.  
The essence of QEK is to find a background whose adjoint action can be identified 
with the spacetime derivative. So, the strategy is {\it to find a set of three matrices 
whose adjoint action can be identified with the derivative on $S^3$}. Such matrices 
were found in \cite{ISTT06,IIST08}. 
As in \S~\ref{sec:4dN=1}, we take the radius of the sphere to be $2/\mu$. 

We introduce matrices $L_i$ which satisfy 
the commutation relation of the $SU(2)$ generators, 
\begin{equation}
 [L_i , L_j ] = i \epsilon_{ijk} L_k . \label{SU(2) commutation}
\end{equation}
Since these matrices cannot be diagonalized simultaneously, 
we embed them in the following block diagonal form; 
\begin{eqnarray}
L_i
=
\left(
\begin{array}{ccccc}
  \ddots & & & & \\
  & L_i^{[j_{s-1/2}]}& & & \\
  & & L_i^{[j_s]} & & \\
  & & & L_i^{[j_{s+1/2}]} & \\
  & & & & \ddots
  \end{array}
\right),  \label{IIST background}
\end{eqnarray}
where $L_i^{[j_s]}$ is a $(2j_s+1)\times (2j_s+1)$ matrix 
which acts on the spin $j_s$ representation. 
The size of the matrix $N$ is 
\begin{eqnarray}
 N=\sum_s (2j_s+1) . 
\end{eqnarray}
We introduce a regularization by restricting 
the representation space to a limited number of $j_s$. 
Furthermore we take the integer $s$ satisfying
\begin{eqnarray}
-\frac{T}{2}\le s\le \frac{T}{2}, 
\end{eqnarray} 
where $T$ is an even integer. 
We introduce another integer $P \ge T/2$ 
and take $j_s$ to be
\begin{eqnarray}
j_s
=
\frac{P+s}{2}.   
\end{eqnarray}
The large $N$ limit is taken in the following way
\begin{eqnarray} 
P
\to\infty, 
\qquad
T\to\infty, 
\qquad
N\to\infty, 
\qquad
T/P\to 0.  
\end{eqnarray} 

By using these matrices we can relate 
a matrix model to a gauge theory on $S^3$ as follows. 
The action of 3d theory can be written as 
\begin{eqnarray}
S_{3d} 
&=&
\frac{N}{\lambda_{3d}}\left(\frac{2}{\mu}\right)^3\int d\Omega_3 Tr
\Biggl(
-
\frac{\mu^2}{4}({\cal L}_iX_j-{\cal L}_jX_i)^2
+
\frac{\mu}{2}({\cal L}_iX_j-{\cal L}_jX_i)[X_i,X_j]
-
\frac{1}{4}[X_i,X_j]^2
\nonumber\\
& &
\qquad\qquad
+
\frac{\mu^2}{2}X_i^2
-
i\mu\epsilon^{ijk}X_iX_jX_k
+
i\mu^2\epsilon^{ijk}X_i({\cal L}_jX_k)
\Biggl),  
\end{eqnarray}
where the derivative ${\cal L}_i$ is defined by (\ref{SU(2) generator on S^3}).  
This can be reproduced from the bosonic three matrix model  
\begin{eqnarray}
S=
\frac{N}{\lambda}\left(
-\frac{1}{4}[X_i,X_j]^2 
-
i\mu\epsilon_{ijk}X^iX^jX^k 
+
\frac{\mu^2}{2}X_i^2
\right),   
\end{eqnarray}
where $\lambda^{-1}=(16\pi^2/\mu^3 NP)\lambda^{-1}_{3d}$ 
($16\pi^2/\mu^3 NP$ 
is a normalization factor analogous to $\Lambda/2\pi$), 
by expanding the action around a classical solution 
\begin{eqnarray}
X_i=-\mu L_i,  
\end{eqnarray}
and identifying ${\cal L}_i$ and $X_i^{(3d)}$ with the counterparts in 0d theory as 
\begin{eqnarray}
{\cal L}_i
\to
[L_i,\ \cdot\ ],
\qquad 
X_i^{(3d)}
\to 
X_i^{(0d)}+\mu L_i, 
\end{eqnarray}
and replacing the trace and the integral by trace, 
\begin{eqnarray}
\left(\frac{2}{\mu}\right)^3\int d\Omega_3 tr
\to
Tr. 
\end{eqnarray}
The UV and IR momentum cutoffs are given by $\mu T$ and $\mu$, respectively, 
and we will take the limit so that 
\begin{eqnarray}
\mu\to 0, 
\qquad
\mu T\to \infty. 
\end{eqnarray}
We also require $\mu^2 P\to\infty$ so that 
spacetime noncommutativity disappears (see \S\ref{sec:NCSYM}). 

Finally we would like to add few remarks. 
First, the background is a classical solution and hence as long as 
it is stable we do not need to quench it. 
Second, when we take the large-$N$ limit fixing the IR momentum cutoff $\mu$, 
to suppress the nonplanar diagrams it is necessary to change the background to 
$-\mu L_i\otimes \textbf{1}_k$ 
and take $k\to\infty$ limit. 
\section{4d ${\cal N}=1$ SYM from matrix quantum mechanics
}\label{EK model for 4d N=1}
\hspace{0.51cm}
In \cite{IIST08}, a regularization of 
4d ${\cal N}=4$ SYM  
on ${\mathbb R}\times S^3$ has been proposed 
by using the BMN matrix model \cite{BMN02}. 
In this section we will generalize this regularization to the case of $\mathcal N=1$ SYM. 
We consider the 4-supercharge matrix quantum mechanics 
given by the dimensional reduction of 4d $\mathcal N=1$ SYM to one dimension. 
We consider its BMN-like deformation \cite{KP06} in order for the matrix model 
to have the matrices (\ref{IIST background}) as a solution. 
Then, applying a similar identification to the one introduced in \cite{IIST08}, 
we obtain a regularization of $\mathcal N=1$ SYM 
on $\mathbb R\times S^3$. 
This regularization keeps all 4 supersymmetries unbroken.
\subsection{BMN-like matrix quantum mechanics}
\hspace{0.51cm}
We start by considering the following matrix quantum mechanics 
\begin{eqnarray}
S_0
=
\frac{N}{\lambda}\int dt Tr\left(
\frac{1}{2}(D_tX_i)^2
+
\frac{1}{4}[X_i,X_j]^2
-
\frac{i}{2}
\bar{\psi}\gamma^0D_t\psi
-
\frac{1}{2}
\bar{\psi}\gamma_i[X_i,\psi]
\right).    
\label{action:4SUSY MQM}
\end{eqnarray} 
Here $X_i\ (i=1,2,3)$ are $N\times N$ traceless hermitian matrices, 
the covariant derivative $D_t$ is defined by $D_t=\partial_t-i[A,\ \cdot\ ]$, 
$\gamma^\mu$ are gamma matrices in four dimensions, and $\psi_\alpha$ are fermionic matrices with 
four-component Majorana spinor index $\alpha$.  
This matrix model is obtained by dimensional reduction of 
4d $\mathcal N=1$ SYM to one dimension, and has 4 supercharges 
which correspond to 4d $\mathcal N=1$ supersymmetry. 
We deform it by adding BMN-like terms \cite{KP06}\footnote{
In \cite{KP06} more general kind of mass deformations to (\ref{action:4SUSY MQM}) 
has been studied systematically. 
}, 
\begin{eqnarray}
S=S_0+S_m, 
\end{eqnarray}
where 
\begin{eqnarray}
S_m
=
\frac{N}{\lambda}\int dt Tr\left( 
\frac{i\beta}{2}\mu
\bar{\psi}\gamma^{123}\psi 
+
i\mu
\epsilon^{ijk}
X_iX_jX_k
-
\frac{\mu^2}{2} X_i^2
\right).
\end{eqnarray}
The additional terms contains a ``mass'' parameter $\mu$. 
We also introduced a constant $\beta$, 
which will be fixed later. 
It is straightforward to see that 
this action is invariant under the SUSY transformations
\begin{eqnarray}
\delta_\epsilon A
&=&
-i\bar{\epsilon}\gamma_0\psi,
\nonumber\\
\delta_\epsilon X_i
&=&
-i\bar{\epsilon}\gamma_i\psi,
\nonumber\\
\delta_\epsilon\psi 
&=&
\left(
(D_tX_i)\gamma^{0i}
-
\frac{i}{2}\gamma^{ij}[X_i,X_j]
+
\frac{1}{2}
\mu X_i\epsilon^{ijk}\gamma^{jk}
\right)\epsilon.  
\label{SUSY transf_matrix model}
\end{eqnarray}
Here $\epsilon$ is a time-dependent parameter
\begin{eqnarray}
\epsilon(t)=e^{-\alpha\mu t\gamma^{0123}}\epsilon_0,  
 \label{epsilon of matrix model}
\end{eqnarray}
where $\alpha$ is a constant which satisfies 
$
\alpha-\beta=1$ , 
and $\epsilon_0$ is a constant Majorana spinor. 
Note that $\epsilon(t)$ satisfies the Majorana condition.   
In fact different choices of $\alpha$ and $\beta$ are related by a time-dependent 
field redefinition \cite{KP06}. As we will see, a specific choice of $\alpha$ and $\beta$ 
is convenient to see the correspondence to 4d ${\cal N}=1$ SYM manifestly.

It turns out that this matrix model has the fuzzy sphere solution. 
To see this we set $\psi=0$ in the equations of motion
\begin{eqnarray}
-[X_j,[X_j,X_i]]
+
\frac{3i}{2}
\mu\epsilon^{ijk}[X_j,X_k]
-
\mu^2 X_i
=0. 
\end{eqnarray}
Hence, 
\begin{eqnarray}
X_i
=
-\mu L_i \label{Background Solution}
\end{eqnarray}
is a classical solution if $L_i$ satisfies the commutation relation 
of the $SU(2)$ generators \eqref{SU(2) commutation}. 
Here we take $L_i$ to be a matrices given in (\ref{IIST background}) in order to obtain the four dimensional theory. 
Furthermore, this solution is invariant under the SUSY transformation.   
\subsection{Correspondence to 4d ${\cal N}=1$ }
\hspace{0.51cm}
Now, we consider the correspondence to 
the $\mathcal N = 1$ SYM on $\mathbb R \times S^3$. 
By expanding around the solution of \eqref{Background Solution}, 
$X_i = - \mu L_i + a_i$, 
the bosonic part of the action becomes 
\begin{eqnarray}
S_{\text{bosonic}} 
&=&
\frac{N}{\lambda}\int dt\ Tr
\Biggl(
\frac{1}{2}
\left(
D_t a_i
-
i\mu [L_i , A_t]
\right)^2
\nonumber\\ 
& &
\qquad\qquad
+
\frac{\mu^2}{4}\left([L_i,a_j]-[L_j,a_i]\right)^2
-
\frac{\mu}{2}\left([L_i, a_j]-[L_j,a_i]\right)[a_i,a_j]
+
\frac{1}{4}[a_i,a_j]^2
\nonumber\\
& &
\qquad\qquad
-
\frac{\mu^2}{2}a_i^2
+
i\mu\epsilon^{ijk}a_ia_ja_k
-
i\mu^2\epsilon^{ijk}a_i[L_j,a_k]
\Biggl). 
\end{eqnarray} 

We can easily see that, by formally replacing 
\begin{eqnarray}
[L_i,\ \cdot\ ] \to {\cal L}_i , 
\qquad
A_t \to A_t(x) , 
\qquad
a_i \to X_i(x) , 
\qquad
Tr \to (2/\mu)^3\int d\Omega_3\ Tr,  
\qquad
\lambda\to\lambda_{4d}, 
\label{Replacement}
\end{eqnarray}
the matrix model and the field theory can be identified.  
Similarly, we can identify the fermionic part of the action.  
The fermionic part of the matrix model can be expressed as 
\begin{eqnarray}
S_{\text{fermionic}}
&=&   
\frac{N}{\lambda}
\left(
-\frac{i}{2}
\right)
\int dt
Tr\left(
\bar{\psi}\gamma^0D_t\psi
+
\bar{\psi}\gamma^i
\left(
i\mu [L_i,\psi]
-
i[a_i,\psi]
\right)
-\beta 
\mu\bar{\psi}\gamma^{123}\psi
\right). 
\nonumber\\
\end{eqnarray}
Hence, by taking the parameters $\alpha$ and $\beta$ of the matrix model as 
$\alpha=\frac{1}{4}$ and $\beta=-\frac{3}{4}$, 
the fermionic part of the SYM and matrix model become manifestly equivalent.   
 
Using the replacement \eqref{Replacement}, 
we can see the correspondence of the SUSY transformations defined in 
\eqref{SUSY Transf on RxS^3} and \eqref{SUSY transf_matrix model}. 
The time dependence of the parameter $\epsilon$ 
is also same for the SYM on $\mathbb R\times S^3$ 
\eqref{epsilon on RxS^3} and the matrix model \eqref{epsilon of matrix model}.  
Furthermore when we take the continuum limit, we have to scale the gauge coupling constant 
appropriately with the UV momentum cutoff. 

Before concluding this section, an important remark is in order.  
In 4d ${\cal N}=1$ SYM, there is a $\left(Z_N\right)^4$-unbroken phase \cite{KUY07}, 
that is volume-independent. (For related works, see \cite{BBCS09,Bringoltz09}).  
One may think that, because of  this volume-independence, 4d ${\cal N}=1$ SYM 
is related to the dimensionally reduced model (\ref{action:4SUSY MQM}). 
However, the situation is not so simple. 
In order for the small volume limit of 4d ${\cal N}=1$ to be described 
by (\ref{action:4SUSY MQM}),   
the $Z_N$ symmetry must be broken \cite{AMMPRW05}.  
If the $Z_N$ is not broken,  the derivative $\partial_\mu$ 
and the commutator $[A_\mu,\ \cdot\ ]$ in the covariant 
derivative give contributions of the same order. 
As a consequence the Kaluza-Klein excited modes and 
the zero modes as well give effective masses of the same order. 
Therefore, even in the small volume limit, we cannot simply truncate  the Kaluza-Klein modes.  
In the original Eguchi-Kawai model, this problem is avoided by using the unitary matrix. 
However it is difficult to keep supersymmetry unbroken with unitary variables. 
This is the reason why 
we have used the technique introduced in \cite{IIST08}.
\subsection{Absence of the sign problem}\label{sec:sign problem}
\hspace{0.51cm}
It is easy to see that 
the regularization described above does not suffer from the notorious sign problem 
after the Wick rotation. 
Firstly the bosonic part of the action is real.  
Therefore it is sufficient to see that the Pfaffian (or the determinant in the Weyl representation) 
of the Dirac operator is free from the sign problem. 
 
When $\beta=0$, the determinant is the same as that of the un-deformed model and 
in the non-lattice regularization method there is no sign problem \cite{HNT07}. 
The proof is a straightforward generalization of that of the zero-dimensional model \cite{AABHN00_1}. 
For the proof, the Weyl representation is more convenient. First let us briefly 
summarize the proof in 0d theory. In the Weyl representation, the Dirac operator $M_{\alpha ij,\beta kl}$, 
which is defined by 
\begin{eqnarray}
\bar{\psi}_{\alpha ji}M_{\alpha ij,\beta kl}\psi_{\beta kl}
=
Tr\bar{\psi}\Gamma^\mu[A_\mu,\psi], 
\end{eqnarray}
reads \footnote{
Strictly speaking we have to project the $U(1)$ part because it gives zero eigenvalue of 
the Dirac operator, but we omit it here just for notational simplicity. 
In the 1d theory at finite temperature it is not necessary. } 
\begin{eqnarray}
M_{\alpha ij,\beta kl}
=
\Gamma^\mu_{\alpha\beta}
\left(
A_{\mu ik}\delta_{jl}
-
A_{\mu lj}\delta_{ik}
\right), 
\end{eqnarray}
where $A_\mu\ (\mu=1,\cdots,4)$ are Hermitian matrices and  
$\psi,\bar{\psi}$ are complex matrices with two-component Weyl indices.   
$\Gamma^\mu$ can be chosen as 
\begin{eqnarray}
\Gamma^\mu=\sigma^\mu\ (\mu=1,2,3), 
\qquad
\Gamma^4 = i\cdot \textbf{1}_2, 
\end{eqnarray}
where $\sigma^\mu$ are the Pauli matrices. 
Let $\varphi_{\alpha,ij}$ to be an eigenvector of $M$ of the eigenvalue $\lambda$. 
Then, noticing that the adjoint operators 
$N_{\mu;ij,kl}
\equiv
A_{\mu ik}\delta_{jl}
-
A_{\mu lj}\delta_{ik}$ 
satisfy 
$N_{\mu;ji,lk}
=
-N_{\mu;ij,kl}^\ast
$
, and $\sigma^2\Gamma^\mu\sigma^2=-(\Gamma^\mu)^\ast$, 
we obtain  
\begin{eqnarray}
\sigma^2_{\alpha\alpha'}
M_{\alpha' ji, \beta' lk}
\sigma^2_{\beta'\beta}
=
\left(M_{\alpha ij, \beta kl}\right)^\ast.  
\end{eqnarray}
Therefore, 
\begin{eqnarray}
M_{\alpha ij,\beta kl}(\sigma^2\varphi^\dagger)_{\beta kl}
&=&
\sigma^2_{\alpha\gamma}
\left(M_{\gamma ji,\beta lk}
\varphi_{\beta lk}\right)^\ast
\nonumber\\
&=&
\lambda^\ast
(\sigma^2\varphi^\dagger)_{\alpha ij},  
\end{eqnarray}
and hence $(\sigma^2\varphi^\dagger)_{\alpha ij}$ is eigenvector of the eigenvalue 
$\lambda^\ast$. Note that they are linearly independent and hence 
the determinant is written as 
\begin{eqnarray}
det M 
=
\prod_i|\lambda_i|^2
\ge 0,  
\end{eqnarray} 
where $i$ is a label for pairs of eigenvalues. 
It is manifestly free from the sign problem. 

Next let us consider the 1d theory. First let us consider the case when $\beta=0$ \cite{HNT07}. 
After the Wick rotation, the fermionic part of the action is 
\begin{eqnarray}
\int dt Tr\left(
\bar{\psi}D_t\psi
-\sum_{\mu=1}^3
\bar{\psi}\sigma^\mu[X_\mu,\psi]
\right).  
\end{eqnarray}
In the momentum cutoff prescription \cite{HNT07}, we compactify the time direction 
with period $1/T$ ($T$ is the temperature) and fix the gauge symmetry so that 
the gauge field becomes static and diagonal, 
\begin{eqnarray}
A(t)=T\cdot diag(\alpha_1,\cdots,\alpha_N), 
\qquad
-\pi\le\alpha_i< \pi. 
\end{eqnarray}
Furthermore we introduce the momentum cutoff $\Lambda\in {\mathbb Z}$ such that  
\begin{eqnarray}
X_\mu(t)
=
\sum_{n=-\Lambda}^\Lambda
\tilde{X}_\mu(n)e^{i\omega nt}, 
\qquad
\psi_\alpha(t)
=
\sum_{r=-\Lambda+1/2}^{\Lambda-1/2}
\tilde{\psi}_\alpha(r)e^{i\omega rt}, 
\qquad
\bar{\psi}_\alpha(t)
=
\sum_{r=-\Lambda+1/2}^{\Lambda-1/2}
\tilde{\bar{\psi}}_\alpha(r)e^{-i\omega rt},  
\end{eqnarray} 
where $\omega=2\pi T$. 
Here $n$ and $r$ run integer and half-integer values, respectively. 
Then the fermionic part becomes 
\begin{eqnarray}
\tilde{\bar{\psi}}_{\alpha ji}(p)
M_{\alpha ij p, \beta kl q}
\tilde{\psi}_{\beta kl}(q)
\end{eqnarray}
where
\begin{eqnarray}
M_{\alpha ij p, \beta kl q}
=
i\left(
p\omega-T(\alpha_i-\alpha_j)
\right)
\delta_{\alpha\beta}\delta_{ik}\delta_{jl}\delta_{pq}
-\sum_{\mu=1}^3\sigma^\mu_{\alpha\beta}
\left(
\tilde{X}_{\mu ik}(p-q)\delta_{jl}
-
\tilde{X}_{\mu lj}(p-q)\delta_{ik}
\right). 
\end{eqnarray}
Note that $M$ has momentum indices $p$ and $q$ in this case. 
Because $\left(\tilde{X}_{\mu ij}(p)\right)^\ast=\tilde{X}_{\mu ji}(-p)$,  
the Dirac operator $M$ satisfies 
\begin{eqnarray}
\sigma^2_{\alpha\alpha'}
M_{\alpha' ji (-p), \beta' lk (-q)}
\sigma^2_{\beta'\beta}
=
\left(M_{\alpha ij p, \beta kl q}\right)^\ast, 
\end{eqnarray}
and hence if $M\varphi=\lambda\varphi$ then 
$M(\sigma^2\varphi^\dagger)=\lambda^\ast(\sigma^2\varphi^\dagger)$. 
Therefore the determinant of $M$ is always equal to or larger than zero.  

For generic values of $\beta$, the Dirac 
operator is shifted by a mass term $\beta\mu\cdot\textbf{1}$ 
and the eigenvalues are shifted simply as 
$\lambda+\beta\mu,\lambda^\ast+\beta\mu=(\lambda+\beta\mu)^\ast$.   
Therefore the determinant is equal to or larger than zero also in this case.  
 
The same discussion is applicable in any dimension, 
as long as the theory is obtained from 4d ${\cal N}=1$, 
and hence with the momentum cutoff the sign does not appear. 
Of course in higher dimensions we need to use the lattice regularization  
and hence the positivity of the determinant is violated. 
However the sign problem is treatable in the following sense, 
at least in less than three-dimensions. 
Suppose that one simulate the model by using the phase quenched action 
$S_{quench}=S_{bosonic}-\log|\det M|$, where $\det M$ is 
the fermion determinant. Then the effect of the phase factor 
$\det M/|\det M|$ can be taken into account by the reweighting as  
$\langle{\cal O}\rangle=\langle{\cal O}\cdot{\rm phase}\rangle_q/\langle{\rm phase}\rangle_q$, 
where $\langle\ \cdot\ \rangle$ and $\langle\ \cdot\ \rangle_q$ represent 
the expectation value of the original and phase quenched models, respectively. 
If fluctuation of the phase becomes large, 
both of the numerator and denominator in r.h.s. become small 
and the numerical error cause fatal problem. 
For lower dimensional theories which is 
obtained from 4d $\mathcal N=1$ SYM, however, 
as one approaches to the continuum limit, 
the phase factor goes close to 1 for most of the configurations; 
see \cite{CW07} for 1d and \cite{Catterall06,Kanamori09} for 2d.    
In such a case, the reweighting method works. 
In this sense the sign problem is treatable. 
It would be nice if similar property can be 
seen in the three- and four-dimensional lattices. 

Although the sign problem is treatable in the sense explained above,  
however, at finite cutoff level small sign effect remains and hence 
in order to calculate the expectation value precisely one has to 
perform the reweighting procedure. 
For that, one has to calculate the fermion determinant. 
One of the advantages of the present method is that the positivity is kept exactly, 
the reweighting is not necessary and hence by using the rational Hybrid Monte-Carlo algorithm 
we do not need to calculate the determinant. This property reduces the computational cost 
drastically.  
\section{4d ${\cal N}=1$ noncommutative SYM  
}\label{sec:NCSYM}
\hspace{0.51cm}
In this section we provide a matrix model formulation of 
4d ${\cal N}=1$ noncommutative SYM. 
First we explain how gauge theories on noncommutative space are obtained 
from the large-$N$ matrix models \cite{AIIKKT99}. Then we discuss the finite-$N$ regularization. 

Let us start with a bosonic $D$-matrix model 
\begin{eqnarray}
S=-\frac{1}{4g^2}\sum_{\mu\neq\nu}
         Tr\left[X_\mu,X_\nu\right]^2. 
\label{IKKT}
\end{eqnarray} 
The model has a classical solution\footnote{
This solution cannot be realized at finite $N$. 
} 
\begin{eqnarray}
X_\mu^{(0)}=\hat{p}_\mu\ (\mu=1,\cdots,d), 
\qquad
X_\mu^{(0)}=0\ (\mu=d+1,\cdots,D), 
\qquad
[\hat{p}_\mu,\hat{p}_\nu]
=
i\theta_{\mu\nu}\cdot\textbf{1}_N,     
\end{eqnarray}
where $\textbf{1}_N$ is an $N\times N$ unit matrix. 
By expanding the action (\ref{IKKT}) around it 
we obtain the $U(1)$ noncommutative Yang-Mills theory (NCYM) 
on the fuzzy space ${\mathbb R}^d$ with $(D-d)$ scalar fields. 
The construction goes as follows: let us define the ``noncommutative coordinate''
$\hat{x}^\mu=\left(\theta^{-1}\right)^{\mu\nu}\hat{p}_\nu$
they satisfy
\begin{eqnarray}
  [\hat{x}^\mu,\hat{x}^\nu]=-i(\theta^{-1})^{\mu\nu}\cdot\textbf{1}_N
\end{eqnarray}
this commutation relation is the same as for 
the coordinates on the fuzzy space ${\mathbb R}^d$ with noncommutativity parameter
$\theta$, and as a consequence the functions of $\hat{x}$ can be mapped to 
functions on the fuzzy space ${\mathbb R}^d$. More precisely, we have the
following mapping rule:  
\begin{eqnarray}
  \begin{array}{ccc}
    f(\hat{x})=\sum_k\tilde{f}(k)e^{ik\hat{x}}
    &\leftrightarrow&
    f(x)=\sum_k\tilde{f}(k)e^{ikx}, 
    \\
    f(\hat{x})g(\hat{x})
    &\leftrightarrow&
    f(x)\star g(x), \\
    i[\hat{p}_\mu,\ \cdot\ ]
    &\leftrightarrow&
    \partial_\mu, \\
    Tr
    &\leftrightarrow&
    \frac{\sqrt{\det\theta}}{4\pi^2}\int d^dx,  
  \end{array}
\end{eqnarray}
where $\star$ represents the Moyal product, 
\begin{eqnarray}
  f(x)\star g(x)
  =
  f(x)\exp\left(-\frac{i}{2}
    \overset{\leftarrow}{\partial}_\mu
    (\theta^{-1})^{\mu\nu}
    \overset{\rightarrow}{\partial}_\nu
  \right) g(x) . 
\end{eqnarray}
Using this prescription we obtain $U(1)$ NCYM with coupling constant 
\begin{equation}
g_{NC}^2=4\pi^2g^2/\sqrt{\det\theta}.
\end{equation}
Similarly, by taking the background to be 
$A_\mu^{(0)}=\hat{p}_\mu\otimes\textbf{1}_k\ (\mu=1,\cdots,d)$ we obtain 
$U(k)$ NCYM. The UV cutoff is $\Lambda\sim\left((N/k)\sqrt{\det\theta}\right)^{1/d}$, 
and $g_{NC}$ should be renormalized appropriately.

\begin{figure}[tbp]
\begin{center}
\scalebox{0.3}{
\rotatebox{0}{
\includegraphics*{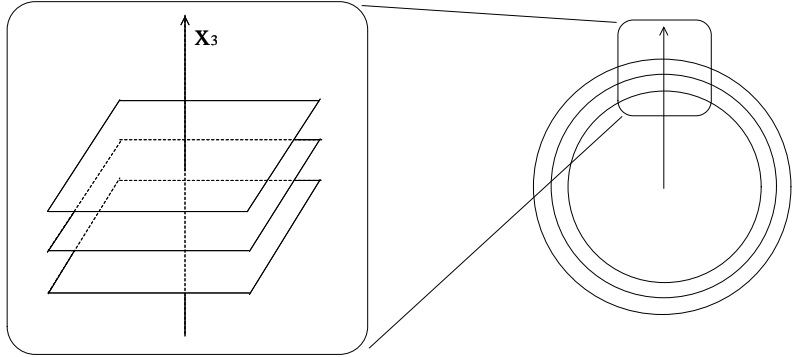}
}}
\caption{Fuzzy planes extending to $(x_1,x_2)$-direction placed along $x_3$ (left) 
can be obtained by zooming in the north pole of a set of concentric fuzzy spheres (right).   
}\label{fig:FuzzyPlanes}
\end{center}
\end{figure}

Next let us combine the above technique with the Eguchi-Kawai prescription. 
We consider the three matrix model, $D=3$, and take the background to be 
\begin{eqnarray}
X_1^{(0)}=\hat{p}_{n_1}\otimes\textbf{1}_{n_2}, 
\qquad
X_2^{(0)}= \hat{q}_{n_1}\otimes\textbf{1}_{n_2},
\qquad
X_3^{(0)}= \textbf{1}_{n_1}\otimes diag(p_1,\cdots,p_{n_2}), 
\label{multi fuzzy plane}
\end{eqnarray}
where
\begin{eqnarray}
[\hat{p}_{n_1},\hat{q}_{n_1}]=-i\theta\cdot\textbf{1}_{n_1}  
\end{eqnarray}  
and 
\begin{eqnarray}
p_k=\frac{\Lambda}{n_2}\left(
k-\frac{n_2}{2}
\right). 
\end{eqnarray}
Intuitively, the fuzzy planes extending into the $(x_1,x_2)$-direction are located
at each value of $x_3$ (Fig.~\ref{fig:FuzzyPlanes}, left). 
Then, in the large-$N$ limit ($n_1,n_2\to\infty$)  
we obtain two noncommutative directions using the previous construction
and one ordinary (commutative) direction by the Eguchi-Kawai prescription. 
Note that the gauge group is $U(\infty)$. 

To realize this configuration at finite-$N$, 
we can use the background (\ref{IIST background}). 
First let us consider the single fuzzy sphere of spin $j$, described by $(2j+1)\times (2j+1)$ matrices.  
By zooming in the north pole, i.e. 
by looking only close to $L_3 = j$, we have 
\begin{eqnarray}
[\mu L_1,\mu L_2]
\simeq
-i\mu^2 j. 
\end{eqnarray}
Hence, we can identify $\mu L_1$ and $\mu L_2$ to be $\hat{p}$ and $\hat{q}$ 
with the noncommutative parameter 
\begin{eqnarray}
\theta = \mu^2 j. 
\end{eqnarray}
The tangent space at the north pole, which is identified with the noncommutative plane, is located at 
\begin{eqnarray}
x_3=\mu j. 
\end{eqnarray}
In order to obtain the three dimensional noncommutative flat space 
described in (\ref{multi fuzzy plane}) we must consider the tangent spaces 
for the whole family of concentric fuzzy spheres defined in (\ref{IIST background}). 
These planes must be required to have the same noncommutative parameter $\theta$, 
but will be placed at different positions $x_3 = \mu j_s$. 
In this description, the distance between the neighboring spheres $\mu$ turns into the infrared momentum 
cutoff along the commutative direction.  
Therefore, we have to take the following limit
\begin{eqnarray}
 \mu^2 j_s \to \theta,
\qquad  
 \mu \to 0. 
\end{eqnarray}

Let us now consider how to impose this limit 
in the background \eqref{IIST background}. 
We first define the infinitesimal parameter $\epsilon$ as $\mu = \epsilon$. 
Then, $j_0 = P$ will behave as 
\begin{equation}
 P \sim \epsilon^{-2} . 
\end{equation}
In order for the noncommutative parameter to coincide, 
the maximum and minimum of $\mu^2 j_s$ must go to same value
\begin{equation}
 \mu^2 \left(P\pm\frac{T}{2}\right) \to \mu^2 P.
\end{equation}
Hence, $T$ must scale as 
\begin{equation}
 T \sim \epsilon^\gamma , 
\end{equation}
with $\gamma > - 2$. 
We must also send the distance $\mu T$ between the first and last tangent planes,
which corresponds to the UV cutoff in the Eguchi-Kawai reduction, 
to infinity, 
\begin{equation}
 \Lambda \sim \mu T \to \infty . 
\end{equation}
From the previous relation we obtain $\gamma < -1$.  
In summary, by sending $T$ and $P$ to infinity 
in the following way: 
\begin{align}
 \mu &\sim \epsilon , &
 P &\sim \epsilon^{-2} , & 
 T &\sim \epsilon^\gamma , &
 \epsilon &\to 0, 
\end{align}
where $-2 < \gamma < -1$,  
we obtain the 3d (Euclidean) noncommutative space from three matrices. 

Similarly, we can obtain $(1+3)$-d ${\cal N}=1$ noncommutative super Yang-Mills
by applying this procedure to the BMN-like matrix model.

\subsection{Stability of the background}
\hspace{0.51cm}
For the bosonic models, noncommutative backgrounds which lead to the ordinary 
flat noncommutative space are unstable \cite{BNSV06,AHHI07,AHH08}. 
Such an instability was originally argued in \cite{VanRaamsdonk01} 
from the point of view of UV/IR mixing \cite{MRS99}. 
The argument in \cite{AHHI07,AHH08} can be applied to the present case as well and 
hence the 4d bosonic NCYM cannot be constructed in this way, as also expected 
from the NCYM calculation \cite{VanRaamsdonk01}\footnote{
Recently it has been claimed that by using Eguchi-Kawai model with 
double trace deformation \cite{UY08} 4d bosonic NCYM can be defined 
\cite{Unsal08}. If so then the deformation should eliminate UV/IR mixing somehow. 
It would be interesting to study this point in detail. 
}.
 
In supersymmetric models this type of instability does not seem to exist. 
Given that the concentric fuzzy sphere background is supersymmetric, 
we can expect it to be stable in the limit discussed above 
at least at zero-temperature. 
It would be interesting to study the stability explicitly 
by using the Monte-Carlo simulation\footnote{
The stability of fuzzy sphere in zero-dimensional supersymmetric matrix model 
has been studied by using Monte-Carlo simulation in \cite{AANN05}. 
}. Given that  noncommutative spaces in bosonic models are unstable, as mentioned previously, we expect the existence of a critical temperature above which the noncommutative space 
destabilizes.  

\section{Conclusion and Discussions}
\hspace{0.51cm}
In this article we have introduced a BMN-like supersymmetric deformation 
of the 4-supercharge matrix quantum mechanics which can be obtained from 
4d ${\cal N}=1$ SYM through dimensional reduction. 
By using it, we provided nonperturbative formulation of 
4d ${\cal N}=1$ planar SYM and 4d ${\cal N}=1$ noncommutative SYM. 
These models can be studied numerically by using the non-lattice  
simulation techniques of supersymmetric matrix quantum mechanics 
\cite{HNT07}\footnote{
In one dimension lattice simulation also works, that is, 
supersymmetry is restored in the continuum limit \cite{CW07}.  
}. 
An important application which we have in mind is the analysis 
of the finite temperature phase structure of ${\cal N}=1$ SYM. 
In theories that have a gravity dual, as ${\cal N}=4$ SYM, it is possible to 
infer the existence of a deconfinement phase transition at strong coupling.  
This is just the usual transition from pure $AdS$ to a black hole 
that takes place on the gravity side. Using the formulation of \cite{IIST08}, 
we may reproduce the transition from gauge theory. 
In the case of  ${\cal N}=1$ SYM the gravity dual is not known and 
as a consequence it is difficult to study the phase structure of 
the theory analytically. 
Our nonperturbative formulation allow us to study 
the finite temperature theory numerically and 
to investigate the presence of a deconfinement phase transition.   
However a possible subtlety is that the background (\ref{IIST background}),  
which is necessary for the Eguchi-Kawai equivalence, may be destabilized 
before the phase transition takes place. Also, it is important to confirm that 
the Eguchi-Kawai construction works at strong coupling. For that, 
it is important to compare with the lattice calculation.  

Another interesting direction is to use the formulation to obtain 
insights into nonsupersymmetric theories.
In \cite{ASV03}, a large-$N$ nonsupersymmetric 
gauge theory with a quark in the two-index representation has been discussed. 
This model is interesting because it can be regarded as a certain large-$N$ 
limit of one-flavor QCD, in the sense that it reduces to the ordinary one at $N=3$. 
It was found that this theory can be embedded into 4d ${\cal N}=1$ SYM, 
and a class of interesting quantities such as the fermion condensate take the same value 
as in the counterpart in SYM at large-$N$. 
Our formulation would be used to obtain insights into the nonsupersymmetric 
theory through this equivalence. 
In this context, it is interesting to study the nonsupersymmetric theory itself, 
not necessarily with one flavor, by using the Eguchi-Kawai equivalence. 
For such systems, two kinds of the large-$N$ reductions 
-- 
the one on $S^3$ \cite{IIST08} discussed in this paper and 
the usual one with unitary variables studied in \cite{Bringoltz09}  
--
are applicable. 
In this case 
a possible disadvantage of the former formulation is that it is necessary to put many fuzzy spheres 
on top of each other in order to stabilize the background against the repulsive 
force coming from many fermions. 
It would be nice if we could study the system by using the Eguchi-Kawai equivalence. 

Another interesting direction is to formulate large-$N$ field theories 
on more complicated spacetime using the Eguchi-Kawai equivalence. 
For example, from the discussion in \S~\ref{sec:3d example}, one can easily see that, 
by restricting the spin of $SU(2)$ generators in \eqref{IIST background} 
to be integer values, one obtains super Yang-Mills 
theory on ${\mathbb R}P^3$. 
By noticing the similarity of the ordinary QEK and Taylor's T-dual prescription 
for matrix models \cite{Taylor96}, a generalization of Taylor's work to nontrivial 
manifolds \cite{T-dual} may lead to other examples of 
Eguchi-Kawai construction for more complicated spaces. (Note that the original 
construction \cite{IIST08} has been obtained in this way). 
Generalization to 4d ${\cal N}=2$ is also possible.  
For that purpose, we have to introduce a mass deformation 
which preserves eight supercharges. 
Such a deformation has been discussed in \cite{KP06}. 
 
Furthermore the BMN-like matrix model has several applications. 
For example, it could be used to numerically confirm the existence of 
the conjectured ``commuting matrix phase'' in the BMN matrix model proposed in \cite{BHH08,AHHS09}. 
(Originally such a phase has been conjectured in 4d ${\cal N}=4$ SYM on 
${\mathbb R}\times S^3$ \cite{Berenstein05}). 
Confirming the existence of this phase is important because 
it would enable us to study ${\cal N}=4$ SYM in the strong coupling regime.
This model can also be used to study the stability of the fuzzy sphere solution in 
a supersymmetric setup\footnote{
Fuzzy sphere stability in the bosonic matrix quantum mechanics 
has been studied numerically in \cite{KNT07}.  
}. This will provide further intuition about the dynamic of the fuzzy sphere solution
for the BMN matrix model.

\section*{Acknowledgements}
\hspace{0.51cm}
The authors would like to thank B.~Bringoltz, G.~Ishiki, 
T.~Tsukioka, M.~\"{U}nsal 
and L.~Yaffe for discussions, 
and V.~Sacksteder for useful comments.

\appendix 
\section{${\cal N}=1$ SYM on $\mathbb R \times S^3$}\label{sec:4dN=1}
\hspace{0.51cm}
In this section, we write down the action for 
$\mathcal N = 1$ SYM 
on $\mathbb R \times S^3$ in a form convenient for our purpose \cite{IIST08} 
(a detailed discussion of SYM on curved spaces can be found in \cite{Blau00} ).
We take the radius of the sphere to be $2/\mu$.
The action of $U(N)$ SYM is given by 
\begin{eqnarray}
S
=
-\frac{N}{\lambda_{4d}}\int dt\int_{S^3} d^3x \sqrt{-g(x)}
Tr\left(
\frac{1}{4}F_{ab}^2 
+
\frac{i}{2}\bar{\psi}\gamma^aD_a\psi
\right) ,   
\end{eqnarray}
where $\lambda_{4d}$ is the 't Hooft coupling constant, $g_{\mu\nu}(x)$ 
is the metric and $g(x)$ is its determinant. 
The field strength is 
\begin{eqnarray}
F_{\mu\nu}
=
\partial_\mu A_\mu
-
\partial_\nu A_\mu
-
i[A_\mu,A_\nu]   
\end{eqnarray}
and
$D_a$ is the covariant derivative defined by
\begin{equation}
 D_a \psi = \partial_a \psi - i [A_a,\psi] + 
  \frac{1}{4}\omega_{abc}\gamma^{bc} \psi
\end{equation}
for adjoint fermions. The Greek indices $\mu$, $\nu$ refer to 
the Einstein frame and the Latin indices to the local Lorentz frame. 

The sphere part of this geometry has the group structure of $SU(2)$. 
Given this group structure, 
there exists a right-invariant 1-form $dg g^{-1}$ 
and dual Killing vectors ${\cal L}_i$, 
satisfying the commutation relation 
\begin{eqnarray}
[{\cal L}_i,{\cal L}_j]=i\epsilon_{ijk}{\cal L}_k. 
\end{eqnarray}
Using the coordinates $(\theta,\psi,\varphi)$ defined by 
$g=e^{-i\varphi\sigma_3/2}e^{-i\theta\sigma_2/2}e^{-i\psi\sigma_3/2}$, 
the vielbein $E^i$ can be expressed as 
\begin{eqnarray}
 E^1 
  &=& 
  \frac{1}{\mu}
  \left(
   -\sin\varphi d\theta + \sin\theta \cos\varphi d\psi 
  \right) , \nonumber\\
 E^2 
  &=& 
  \frac{1}{\mu}
  \left(
   \cos\varphi d\theta + \sin\theta \sin\varphi d\psi    
  \right) , \nonumber\\
 E^3 
  &=& 
  \frac{1}{\mu}
  \left(
   d\varphi + \cos\theta d\psi 
  \right) . 
\end{eqnarray}
In these coordinates the metric is given by 
\begin{equation}
 ds^2 = \frac{1}{\mu^2}
  \left[
   d\theta^2 + \sin^2\theta\,d\varphi^2 
   + \left(d\psi + \cos\theta\,d\varphi\right)^2
  \right] . 
\end{equation}
The spin connection $\omega_{abc}$ can be read off 
from the Maurer-Cartan equation, 
\begin{eqnarray}
 dE^i - \omega^i_{\ jk} E^j \wedge E^k &=& 
 0 , \\
 \omega_{ijk} &=& \frac{\mu}{2}\epsilon_{ijk} . 
\end{eqnarray}
and the Killing vectors are given by 
\begin{eqnarray}
{\cal L}_i=-\frac{i}{\mu}E_i^M\partial_M, 
\end{eqnarray}
where
\begin{eqnarray}
{\cal L}_1
&=&
-i\left(
-\sin\varphi\partial_\theta
-
\cot\theta\cos\varphi\partial_\varphi
+
\frac{\cos\varphi}{\sin\theta}\partial_\psi
\right), 
\nonumber\\
{\cal L}_2
&=&
-i\left(
\cos\varphi\partial_\theta
-
\cot\theta\sin\varphi\partial_\varphi
+
\frac{\sin\varphi}{\sin\theta}\partial_\psi
\right), 
\nonumber\\
{\cal L}_3
&=&
-i\partial_\varphi.
\label{SU(2) generator on S^3} 
\end{eqnarray} 
The Killing vectors represent a  complete basis for the tangent space on $S^3$. 
Furthermore given that  the vielbeins are defined everywhere on $S^3$, 
the indices $i$ can be used as a label for
the vector fields and 1-forms.%
\footnote{
This property is necessary in order to identify this index with 
the one in the matrix model \cite{HKK05}. 
} 

By using the Killing vectors ${\cal L}_i$, 
the bosonic part of the action can be rewritten as \cite{IIST08}
\begin{eqnarray}
S_{\text{boson}} 
&=&
\left(
\frac{2}{\mu}
\right)^3
\frac{N}{\lambda_{4d}}\int dt\int d\Omega_3 Tr
\Biggl(
\frac{1}{2}
\left(
D_t X_i
-
\mu{\cal L}_iA_t
\right)^2
\nonumber\\ 
& &
\qquad\qquad
+
\frac{\mu^2}{4}({\cal L}_iX_j-{\cal L}_jX_i)^2
-
\frac{\mu}{2}({\cal L}_iX_j-{\cal L}_jX_i)[X_i,X_j]
+
\frac{1}{4}[X_i,X_j]^2
\nonumber\\
& &
\qquad\qquad
-
\frac{\mu^2}{2}X_i^2
+
i\mu\epsilon^{ijk}X_iX_jX_k
-
i\mu^2\epsilon^{ijk}X_i({\cal L}_jX_k)
\Biggl) , 
\end{eqnarray} 
where $X_i$ is defined in such a way that the 1-form of the gauge field 
on $S^3$ take the form $A = X_i E^i$, and 
$d\Omega_3$ is the volume form of the unit three-sphere.  
The fermionic part can be expressed as 
\begin{eqnarray}
S_{\text{fermion}} 
&=&  
\frac{N}{\lambda_{4d}}
\left(
-\frac{i}{2}
\right)
\left(
\frac{2}{\mu}
\right)^3
\int dt\int d\Omega_3
Tr\left(
\bar{\psi}\gamma^0D_0\psi
+
\bar{\psi}\gamma^i
\left(
i\mu{\cal L}_i\psi
-
i[X_i,\psi]
\right)
+
\frac{3}{4}
\mu\bar{\psi}\gamma^{123}\psi
\right). \nonumber\\
\end{eqnarray}  

The SUSY transformations of ${\cal N}=1$ SYM on  
curved background 
are given by 
\begin{eqnarray}
\delta A_a
&=&
i\bar{\psi}\gamma_a\epsilon, 
\\
\delta\psi
&=&
\frac{1}{2}F^{ab}\gamma_{ab}\epsilon. \label{SUSY Transf on RxS^3}
\end{eqnarray} 
The parameter $\epsilon$ is related to 
the isometry of the geometry. 
In the case of $\mathbb R\times S^3$, 
$\epsilon$ must satisfy 
\begin{eqnarray}
\nabla_a\epsilon
=
\frac{\mu}{4}\gamma_a\gamma^{123}\epsilon, 
\label{killing eq} 
\end{eqnarray}
which corresponds to the Killing spinor equation in supergravity. 
Actually, given two spinors $\epsilon$ and $\zeta$ satisfying
\eqref{killing eq}, $\bar\zeta\gamma^\mu\epsilon$ is a Killing vector. 
The solution to (\ref{killing eq}) is given by
\begin{eqnarray}
\epsilon=e^{-\frac{1}{4}\mu t\gamma^{0123}}\epsilon_0,  
 \label{epsilon on RxS^3}
\end{eqnarray}
where $\epsilon_0$ is a constant.



\begin{thebibliography}{3}

 
\bibitem{CKU09}
  S.~Catterall, D.~B.~Kaplan and M.~Unsal,
  {\it ``Exact lattice supersymmetry,''} 
  arXiv:0903.4881 [hep-lat].


\bibitem{EK82}
  T.~Eguchi and H.~Kawai,
  ``Reduction Of Dynamical Degrees Of Freedom In The Large N Gauge Theory,''
  Phys.\ Rev.\ Lett.\  {\bf 48} (1982) 1063.

\bibitem{BFSS96}
 T.~Banks, W.~Fischler, S.~H.~Shenker and L.~Susskind,
``M theory as a matrix model: A conjecture,''
 Phys.\ Rev.\  D {\bf 55} (1997) 5112;
 hep-th/9610043.

\bibitem{IKKT96}
 N.~Ishibashi, H.~Kawai, Y.~Kitazawa and A.~Tsuchiya,
 ``A large-N reduced model as superstring,''
 Nucl.\ Phys.\  B {\bf 498} (1997) 467;
 hep-th/9612115.
 
\bibitem{DVV97}
  R.~Dijkgraaf, E.~P.~Verlinde and H.~L.~Verlinde,
  ``Matrix string theory,''
  Nucl.\ Phys.\  B {\bf 500}, 43 (1997); 
  hep-th/9703030.
  
   
\bibitem{Maldacena97}
  J.~M.~Maldacena,
 ``The large N limit of superconformal field theories and supergravity,'' 
  Adv.\ Theor.\ Math.\ Phys.\  {\bf 2} (1998) 231 
  [Int.\ J.\ Theor.\ Phys.\  {\bf 38} (1999) 1113]; 
  hep-th/9711200.  

\bibitem{IMSY98}
  N.~Itzhaki, J.~M.~Maldacena, J.~Sonnenschein and S.~Yankielowicz,
  {\it ``Supergravity and the large N limit of theories with sixteen
   supercharges,''} 
  Phys.\ Rev.\  D {\bf 58}, 046004 (1998)
  [arXiv:hep-th/9802042].


\bibitem{HNT07}
  M.~Hanada, J.~Nishimura and S.~Takeuchi,
  {\it ``Non-lattice simulation for supersymmetric gauge theories in one
   dimension,''} 
   Phys.\ Rev.\ Lett.\  {\bf 99} (2007) 161602; 
  arXiv:0706.1647 [hep-lat].

  K.~N.~Anagnostopoulos, M.~Hanada, J.~Nishimura and S.~Takeuchi,
  {\it ``Monte Carlo studies of supersymmetric matrix quantum mechanics with sixteen
   supercharges at finite temperature,''} 
   Phys.\ Rev.\ Lett.\  {\bf 100} (2008) 021601; 
  arXiv:0707.4454 [hep-th].
 
\bibitem{CW07}
  S.~Catterall and T.~Wiseman,
  {\it ``Towards lattice simulation of the gauge theory duals to black holes and hot
   strings,''} 
  JHEP {\bf 0712} (2007) 104; 
  arXiv:0706.3518 [hep-lat]. 
  
  S.~Catterall and T.~Wiseman,
  {\it ``Black hole thermodynamics from simulations of lattice Yang-Mills theory,''} 
  Phys.\ Rev.\  D {\bf 78}, 041502 (2008); 
  arXiv:0803.4273 [hep-th]. 
 
\bibitem{HNT08}
  M.~Hanada, A.~Miwa, J.~Nishimura and S.~Takeuchi,
  ``Schwarzschild radius from Monte Carlo calculation of the Wilson loop in
  supersymmetric matrix quantum mechanics,''
  arXiv:0811.2081 [hep-th].

  M.~Hanada, Y.~Hyakutake, J.~Nishimura and S.~Takeuchi,
  ``Higher derivative corrections to black hole thermodynamics from
  supersymmetric matrix quantum mechanics,''
  arXiv:0811.3102 [hep-th].

  
\bibitem{KNS98}
  W.~Krauth, H.~Nicolai and M.~Staudacher,
  ``Monte Carlo approach to M-theory,''
  Phys.\ Lett.\  B {\bf 431}, 31 (1998)
  [arXiv:hep-th/9803117].
   

\bibitem{AABHN00_1}
  J.~Ambjorn, K.~N.~Anagnostopoulos, W.~Bietenholz, T.~Hotta and J.~Nishimura,
  ``Large N dynamics of dimensionally reduced 4D SU(N) super Yang-Mills
  theory,''
  JHEP {\bf 0007}, 013 (2000)
  [arXiv:hep-th/0003208].


\bibitem{AABHN00_2}
  J.~Ambjorn, K.~N.~Anagnostopoulos, W.~Bietenholz, T.~Hotta and J.~Nishimura,
  ``Monte Carlo studies of the IIB matrix model at large N,''
  JHEP {\bf 0007} (2000) 011
  [arXiv:hep-th/0005147].
 
\bibitem{mean_field}
  J.~Nishimura and F.~Sugino,
  ``Dynamical generation of four-dimensional space-time in the IIB matrix
  model,''
  JHEP {\bf 0205}, 001 (2002)
  [arXiv:hep-th/0111102].

\bibitem{KKN98}
  For example, see the following papers:

  D.~J.~Gross and E.~Witten,
  ``Possible Third Order Phase Transition In The Large N Lattice Gauge
  Theory,''
  Phys.\ Rev.\  D {\bf 21}, 446 (1980).

  D.~Karabali, C.~J.~Kim and V.~P.~Nair,
  ``On the vacuum wave function and string tension of Yang-Mills theories  in
  (2+1) dimensions,''
  Phys.\ Lett.\  B {\bf 434}, 103 (1998); 
  hep-th/9804132.

\bibitem{Teper08}
  M.~Teper,
  ``Large N,''
  arXiv:0812.0085 [hep-lat].

\bibitem{IIST08}
  T.~Ishii, G.~Ishiki, S.~Shimasaki and A.~Tsuchiya,
  ``N=4 Super Yang-Mills from the Plane Wave Matrix Model,''
  Phys.\ Rev.\  D {\bf 78}, 106001 (2008)
  [arXiv:0807.2352 [hep-th]]. 
 
\bibitem{BMN02}
  D.~E.~Berenstein, J.~M.~Maldacena and H.~S.~Nastase,
  ``Strings in flat space and pp waves from N = 4 super Yang Mills,''
  JHEP {\bf 0204}, 013 (2002)
  [arXiv:hep-th/0202021].

\bibitem{4dN=1LatticeSimulation}
  J.~Giedt, R.~Brower, S.~Catterall, G.~T.~Fleming and P.~Vranas,
  ``Lattice super-Yang-Mills using domain wall fermions in the chiral limit,''
  Phys.\ Rev.\  D {\bf 79}, 025015 (2009)
  [arXiv:0810.5746 [hep-lat]].

  M.~G.~Endres,
  ``Dynamical simulation of N=1 supersymmetric Yang-Mills theory with domain
  wall fermions,''
  arXiv:0902.4267 [hep-lat].

\bibitem{IKNT08}
  G.~Ishiki, S.~W.~Kim, J.~Nishimura and A.~Tsuchiya,
  ``Deconfinement phase transition in ${\cal N}=4$ super Yang-Mills theory on $R\times S^3$ from
  supersymmetric matrix quantum mechanics,''
  Phys.\ Rev.\ Lett.\  {\bf 102}, 111601 (2009)
  [arXiv:0810.2884 [hep-th]].

  Y.~Kitazawa and K.~Matsumoto,
  ``N=4 Supersymmetric Yang-Mills on $S^3$ in Plane Wave Matrix Model at Finite Temperature,'' 
  Phys.\ Rev.\ D {\bf 79}, 065003 (2009)
  [arXiv:0811.0529 [hep-th]].
 
\bibitem{AIIKKT99}
 H.~Aoki, N.~Ishibashi, S.~Iso, H.~Kawai, Y.~Kitazawa and T.~Tada,
  ``Noncommutative Yang-Mills in IIB matrix model,'' 
 Nucl.\ Phys.\  B {\bf 565} (2000) 176;
 hep-th/9908141.

  J.~Ambjorn, Y.~M.~Makeenko, J.~Nishimura and R.~J.~Szabo,
  ``Finite N matrix models of noncommutative gauge theory,''
  JHEP {\bf 9911}, 029 (1999); 
  hep-th/9911041

 M.~Li,
 ``Strings from IIB matrices,'' 
 Nucl.\ Phys.\  B {\bf 499} (1997) 149;
 hep-th/9612222.

 J.~Madore, S.~Schraml, P.~Schupp and J.~Wess,
  ``Gauge theory on noncommutative spaces,''
  Eur.\ Phys.\ J.\  C {\bf 16}, 161 (2000); 
  hep-th/0001203.
 

\bibitem{NN03}
  R.~Narayanan and H.~Neuberger,
  ``Large N reduction in continuum,''
  Phys.\ Rev.\ Lett.\  {\bf 91}, 081601 (2003); 
  hep-lat/0303023.
  
\bibitem{BHN82}
  G.~Bhanot, U.~M.~Heller and H.~Neuberger,
  ``The Quenched Eguchi-Kawai Model,''
  Phys.\ Lett.\  B {\bf 113}, 47 (1982).

\bibitem{Parisi82}  
  G.~Parisi,
  ``A Simple Expression For Planar Field Theories,''
  Phys.\ Lett.\  B {\bf 112}, 463 (1982).

  D.~J.~Gross and Y.~Kitazawa,
  ``A Quenched Momentum Prescription For Large N Theories,''
  Nucl.\ Phys.\  B {\bf 206}, 440 (1982).
  
  S.~R.~Das and S.~R.~Wadia,
  ``Translation Invariance And A Reduced Model For Summing Planar Diagrams In
  QCD,''
  Phys.\ Lett.\  B {\bf 117}, 228 (1982)
  [Erratum-ibid.\  B {\bf 121}, 456 (1983)].
   

\bibitem{GAO82}
 A.~Gonzalez-Arroyo and M.~Okawa,
``The Twisted Eguchi-Kawai Model:
 A Reduced Model For Large N Lattice Gauge Theory,'' 
 Phys.\ Rev.\ D {\bf 27} (1983) 2397. 
 
\bibitem{UY08}
  M.~Unsal and L.~G.~Yaffe,
  {\it ``Center-stabilized Yang-Mills theory: confinement and large $N$ volume
   independence,''} 
  arXiv:0803.0344 [hep-th].
  
   
\bibitem{AHHI07}  
  T.~Azeyanagi, M.~Hanada, T.~Hirata and T.~Ishikawa,
   ``Phase structure of twisted Eguchi-Kawai model,''  
  JHEP {\bf 0801} (2008) 025; 
  arXiv:0711.1925 [hep-lat].   

\bibitem{TV06}  
   M.~Teper and H.~Vairinhos,
  ``Symmetry breaking in twisted Eguchi-Kawai models,'' 
   Phys.\ Lett.\  B {\bf 652} (2007) 359; 
  arXiv:hep-th/0612097.

\bibitem{BNSV06}  
   W.~Bietenholz, J.~Nishimura, Y.~Susaki and J.~Volkholz,
 ``A non-perturbative study of 4d U(1) non-commutative gauge
   theory: The fate of one-loop instability,''
 JHEP {\bf 0610} (2006) 042;
 hep-th/0608072.

\bibitem{BS08}
  B.~Bringoltz and S.~R.~Sharpe,
  ``Breakdown of large-N quenched reduction in SU(N) lattice gauge theories,''
  arXiv:0805.2146 [hep-lat].  

\bibitem{AHH08}
  T.~Azeyanagi, M.~Hanada and T.~Hirata,
  ``On Matrix Model Formulations of Noncommutative Yang-Mills Theories,''
  Phys.\ Rev.\  D {\bf 78}, 105017 (2008)
  [arXiv:0806.3252 [hep-th]].


\bibitem{KS07}
  H.~Kawai and M.~Sato,
  ``Perturbative Vacua from IIB Matrix Model,''
  Phys.\ Lett.\  B {\bf 659}, 712 (2008)
  [arXiv:0708.1732 [hep-th]].

\bibitem{ISTT06}
  G.~Ishiki, S.~Shimasaki, Y.~Takayama and A.~Tsuchiya,
   ``Embedding of theories with $SU(2|4)$ symmetry into the plane wave matrix
   model,''
  JHEP {\bf 0611}, 089 (2006)
  [arXiv:hep-th/0610038].

\bibitem{KP06}
  N.~Kim and J.~H.~Park,
  {\it ``Massive super Yang-Mills quantum mechanics: Classification and the relation
   to supermembrane},''
  Nucl.\ Phys.\  B {\bf 759}, 249 (2006)
  [arXiv:hep-th/0607005].

\bibitem{KUY07}
  P.~Kovtun, M.~Unsal and L.~G.~Yaffe,
  ``Volume independence in large N(c) QCD-like gauge theories,''
  JHEP {\bf 0706}, 019 (2007)
  [arXiv:hep-th/0702021].

\bibitem{BBCS09}
  P.~F.~Bedaque, M.~I.~Buchoff, A.~Cherman and R.~P.~Springer,
  ``Can fermions save large N dimensional reduction?,''
  arXiv:0904.0277 [hep-th].

\bibitem{Bringoltz09}
  B.~Bringoltz,
  ``Large-N volume reduction of lattice QCD with adjoint Wilson fermions at
   weak-coupling,''
  arXiv:0905.2406 [hep-lat].
  
  B.~Bringoltz and S.~R.~Sharpe,
  ``Non-perturbative volume-reduction of large-N QCD with adjoint fermions,''
  arXiv:0906.3538 [hep-lat].


\bibitem{AMMPRW05}
  O.~Aharony, J.~Marsano, S.~Minwalla, K.~Papadodimas, M.~Van Raamsdonk and T.~Wiseman,
  {\it ``The phase structure of low dimensional large N gauge theories on tori,''} 
  JHEP {\bf 0601}, 140 (2006)
  [arXiv:hep-th/0508077].
 

  
\bibitem{Catterall06}
  S.~Catterall,
  ``On the restoration of supersymmetry in twisted two-dimensional lattice
  Yang-Mills theory,''
  JHEP {\bf 0704}, 015 (2007)
  [arXiv:hep-lat/0612008].

\bibitem{Kanamori09}
  I.~Kanamori,
  ``Vacuum energy of two-dimensional N=(2,2) super Yang-Mills theory,''
  Phys.\ Rev.\  D {\bf 79}, 115015 (2009)
  [arXiv:0902.2876 [hep-lat]].


\bibitem{VanRaamsdonk01}
  M.~Van Raamsdonk,
  ``The meaning of infrared singularities in noncommutative gauge theories,'' 
  JHEP {\bf 0111} (2001) 006; 
  hep-th/0110093. 
  
  A.~Armoni and E.~Lopez,
  ``UV/IR mixing via closed strings and tachyonic instabilities,'' 
  Nucl.\ Phys.\  B {\bf 632} (2002) 240; 
  hep-th/0110113.   
 
\bibitem{MRS99}
 S.~Minwalla, M.~Van Raamsdonk and N.~Seiberg,
  ``Noncommutative perturbative dynamics,''
  JHEP {\bf 0002},(2000) 020 ;
  hep-th/9912072.

\bibitem{Unsal08}
  M.~Unsal,
  ``Deformed matrix models, supersymmetric lattice twists and N=1/4
  supersymmetry,''
  arXiv:0809.3216 [hep-lat].
  
\bibitem{AANN05}
  K.~N.~Anagnostopoulos, T.~Azuma, K.~Nagao and J.~Nishimura,
  {\it ``Impact of supersymmetry on the nonperturbative dynamics of fuzzy
   spheres,''} 
  JHEP {\bf 0509} (2005) 046; hep-th/0506062.  


\bibitem{ASV03}
  A.~Armoni, M.~Shifman and G.~Veneziano,
  ``Exact results in non-supersymmetric large N orientifold field theories,''
  Nucl.\ Phys.\  B {\bf 667}, 170 (2003)
  [arXiv:hep-th/0302163].

  A.~Armoni, M.~Shifman and G.~Veneziano,
  ``SUSY relics in one-flavor QCD from a new 1/N expansion,''
  Phys.\ Rev.\ Lett.\  {\bf 91}, 191601 (2003)
  [arXiv:hep-th/0307097].



\bibitem{Taylor96}
    W.~Taylor,
  ``D-brane field theory on compact spaces,''
  Phys.\ Lett.\  B {\bf 394}, 283 (1997)
  [arXiv:hep-th/9611042].

\bibitem{T-dual}  
    T.~Ishii, G.~Ishiki, S.~Shimasaki and A.~Tsuchiya,
  ``T-duality, fiber bundles and matrices,''
  JHEP {\bf 0705}, 014 (2007)
  [arXiv:hep-th/0703021].

  T.~Ishii, G.~Ishiki, K.~Ohta, S.~Shimasaki and A.~Tsuchiya,
  ``On relationships among Chern-Simons theory, BF theory and matrix model,''
  Prog.\ Theor.\ Phys.\  {\bf 119}, 863 (2008)
  [arXiv:0711.4235 [hep-th]].

  T.~Ishii, G.~Ishiki, S.~Shimasaki and A.~Tsuchiya,
  ``Fiber Bundles and Matrix Models,''
  Phys.\ Rev.\  D {\bf 77}, 126015 (2008)
  [arXiv:0802.2782 [hep-th]].


 \bibitem{BHH08}
  D.~E.~Berenstein, M.~Hanada and S.~A.~Hartnoll,
  {\it ``Multi-matrix models and emergent geometry,''} 
  arXiv:0805.4658 [hep-th].   

   
 \bibitem{AHHS09}
   T.~Azeyanagi, M.~Hanada, T.~Hirata and H.~Shimada,
  {\it ``On the shape of a D-brane bound state and its topology change,''} 
  JHEP {\bf 0903}, 121 (2009)
  [arXiv:0901.4073 [hep-th]].

\bibitem{Berenstein05}
  D.~Berenstein,
  ``Large N BPS states and emergent quantum gravity,''
  JHEP {\bf 0601}, 125 (2006)
  [arXiv:hep-th/0507203].

 \bibitem{KNT07}
  N.~Kawahara, J.~Nishimura and S.~Takeuchi,
  {\it ``Exact fuzzy sphere thermodynamics in matrix quantum mechanics,''} 
  JHEP {\bf 0705}, 091 (2007)
  [arXiv:0704.3183 [hep-th]].


  H.~Kawai, S.~Kawamoto, T.~Kuroki, T.~Matsuo and S.~Shinohara,
  ``Mean field approximation of IIB matrix model and emergence of four
  dimensional space-time,''
  Nucl.\ Phys.\  B {\bf 647}, 153 (2002)
  [arXiv:hep-th/0204240].


 
\bibitem{Blau00}
  M.~Blau,
  {\it ``Killing spinors and SYM on curved spaces,''} 
  JHEP {\bf 0011}, 023 (2000)
  [arXiv:hep-th/0005098].
 
\bibitem{HKK05}
  M.~Hanada, H.~Kawai and Y.~Kimura,
  {\it ``Describing curved spaces by matrices,''} 
  Prog.\ Theor.\ Phys.\  {\bf 114}, 1295 (2006)
  [arXiv:hep-th/0508211].
 

\end{thebibliography}
\end{document}